\documentclass[english]{sbrt}
\usepackage[english]{babel}
\usepackage[utf8]{inputenc}

\usepackage{hyperref}
\usepackage{natbib}
\usepackage{graphicx}
\usepackage{booktabs}
\usepackage{multirow}
\usepackage{rotating}
\usepackage{subfig}  % for \subfloat
\usepackage{caption} % in your preamble
\usepackage{array} % in preamble
\usepackage[affil-it]{authblk}

\begin{document}

\title{"Explainable Deep Learning for Cataract Detection in Retinal Images: A Dual-Eye and Knowledge Distillation Approach"}

\author[1*]{MohammadReza Abbaszadeh Bavil Soflaei}
\author[2]{Karim SamadZamini}

\affil[1]{Department of Computer Science, Northern Illinois University, DeKalb, IL, USA}

\affil[2]{%
	Department of Computer Engineering, University College of Nabi Akram, Tariz-Iran}
	
\affil[*]{Corresponding author: Mohammad Reza Abbaszadeh Bavil Soflaei, z2068312@students.niu.edu}

\maketitle

\begin{abstract}
Cataract remains a leading cause of visual impairment worldwide, and early detection from retinal imaging is critical for timely intervention. We present a deep learning pipeline for cataract classification using the Ocular Disease Recognition dataset, containing left and right fundus photographs from 5000 patients. We evaluated CNNs, transformers, lightweight architectures, and knowledge-distilled models. The top-performing model, Swin-Base Transformer, achieved 98.58\% accuracy and an F1-score of 0.9836. A distilled MobileNetV3, trained with Swin-Base knowledge, reached 98.42\% accuracy and a 0.9787 F1-score with greatly reduced computational cost. The proposed dual-eye Siamese variant of the distilled MobileNet, integrating information from both eyes, achieved an accuracy of 98.21\%. Explainability analysis using Grad-CAM demonstrated that the CNNs concentrated on medically significant features, such as lens opacity and central blur. These results show that accurate, interpretable cataract detection is achievable even with lightweight models, supporting potential clinical integration in resource-limited settings.
\end{abstract}

\section{Introduction}

Cataracts are the world’s leading cause of vision loss from disease, affecting more than 94 million people aged 50 and older in 2020 \cite{steinmetz2021causes}. With populations aging, this number is set to rise. Cataract surgery is both effective and affordable, yet millions still go undiagnosed and untreated \cite{Chua2017-ui}, making early detection through accessible screening tools essential. Fundus photography offers a quick, non-invasive, and widely available way to capture images of the retina, revealing signs of cataracts such as blurring, loss of optic disc clarity, and reduced vessel contrast \cite{Xie2023-kb}. Unlike slit-lamp or OCT examinations, it often requires no eye dilation and can be used in community clinics or low-resource settings \cite{Khan2024-pm}. Deep learning (DL) has already proven its ability to match or even surpass expert performance in detecting diseases like diabetic retinopathy \cite{Ting2017-te}, age-related macular degeneration \cite{Leng2023-ws}, and glaucoma \cite{Ling2025-qh,Kermany2018-ti}. Building on these successes, DL is now a promising approach for detecting cataracts from fundus images. In this study, we evaluate a range of architectures—including convolutional neural networks (CNNs) \cite{lecun2015deep}, Vision Transformers (ViTs) \cite{dosovitskiy2021imageworth16x16words}, and lightweight models such as MobileNet \cite{howard2017mobilenetsefficientconvolutionalneural}—for cataract classification. We employ knowledge distillation \cite{hinton2015distillingknowledgeneuralnetwork} to transfer representations from large, high-performing models to compact student networks, to balance accuracy and computational efficiency.We also introduce a novel dual-eye Siamese network \cite{koch2015siamese} that processes left and right fundus images together, leveraging correlations between the eyes to improve performance. To improve model interpretability, we apply Grad-CAM \cite{selvaraju2017grad} for CNNs, enabling comparison of model focus with known clinical markers.

\section{Related Work}

Deep learning has transformed ophthalmic disease detection, notably for diabetic retinopathy (DR) \cite{Atwany2022-qs}, glaucoma, and age-related macular degeneration (AMD) \cite{Deng2022-bx}. CNNs applied to fundus images have achieved expert-level accuracy in DR diagnosis and grading \cite{Tsiknakis2021-oc, Alyoubi2020-mz}. For AMD and glaucoma, deep learning models analyzing OCT and fundus images have demonstrated high sensitivity and specificity, sometimes surpassing human experts \cite{Koseoglu2023-ms, Sheng2022-jd}. Multi-disease AI tools capable of simultaneous DR and AMD detection show promise for integrated screening \cite{Gonz_lez_Gonzalo_2019}.

Despite cataract’s clinical importance, deep learning on fundus images for cataract detection remains less explored. Surveys highlight a gap in fundus-based systems compared to other imaging modalities \cite{zhang2022machinelearningcataractclassification}. Some recent studies developed CNN-based models such as MobileNet-V2 and DenseNet variants, achieving accuracies above 90\% \cite{Elloumi2021-gm, Xie2023-ha, padalia2022cnnlstmcombinationnetworkcataract}. However, literature is sparse on diverse architectures, benchmarking, and lightweight explainable models, especially for resource-limited settings.

Standard CNNs like ResNet and EfficientNet are widely used in medical imaging due to robust feature extraction and scalability \cite{he2015deepresiduallearningimage, tan2020efficientnetrethinkingmodelscaling}. Vision Transformers (ViT, Swin, DeiT) have gained traction for modeling long-range dependencies with strong classification performance \cite{dosovitskiy2021imageworth16x16words, liu2021swintransformerhierarchicalvision, touvron2021trainingdataefficientimagetransformers}. Explainability techniques such as Grad-CAM and attention visualization enhance clinical trust by localizing decision-relevant regions \cite{Selvaraju_2019, dosovitskiy2021imageworth16x16words}. To the best of our knowledge, no previous study has combined knowledge distillation with a dual-eye Siamese architecture for cataract detection, making this a novel fusion of model compression and inter-eye relational learning. For a summarized overview of these studies, see Table~\ref{tab:related_work} in Appendix~A.

\section{Methodology}

This section outlines our approach to automated cataract detection from retinal fundus images, covering dataset preprocessing, the design and training of single-eye and dual-eye models, the application of knowledge distillation, and the use of explainability techniques—each contributing to improved performance and interpretability.

\subsection{Dataset}

We used the publicly available Ocular Disease Intelligent Recognition (ODIR-5K) dataset from Kaggle \cite{larxel2022odir5k}, which includes paired left and right fundus photographs from approximately 5,000 patients (around 10,000 images total). Each record contains high-resolution color images and metadata such as age, gender, and diagnostic labels for multiple ocular diseases. This paired-eye format supports both single-eye and dual-eye learning approaches.

For this study, we focused on the binary classification task of differentiating “Normal” versus “Cataract” eyes by filtering samples accordingly. The resulting dataset is heavily imbalanced, with 2,873 normal images and only 293 cataract images. To manage data loading and preprocessing, we implemented a custom PyTorch dataset class, \texttt{FundusDataset}, which accepts a pandas \texttt{DataFrame} containing image filenames and labels, applying specified transformations during access.

\begin{figure}[hbt]
	\begin{center}
		\setlength{\unitlength}{0.0105in}%
		\includegraphics[width=0.9\linewidth]{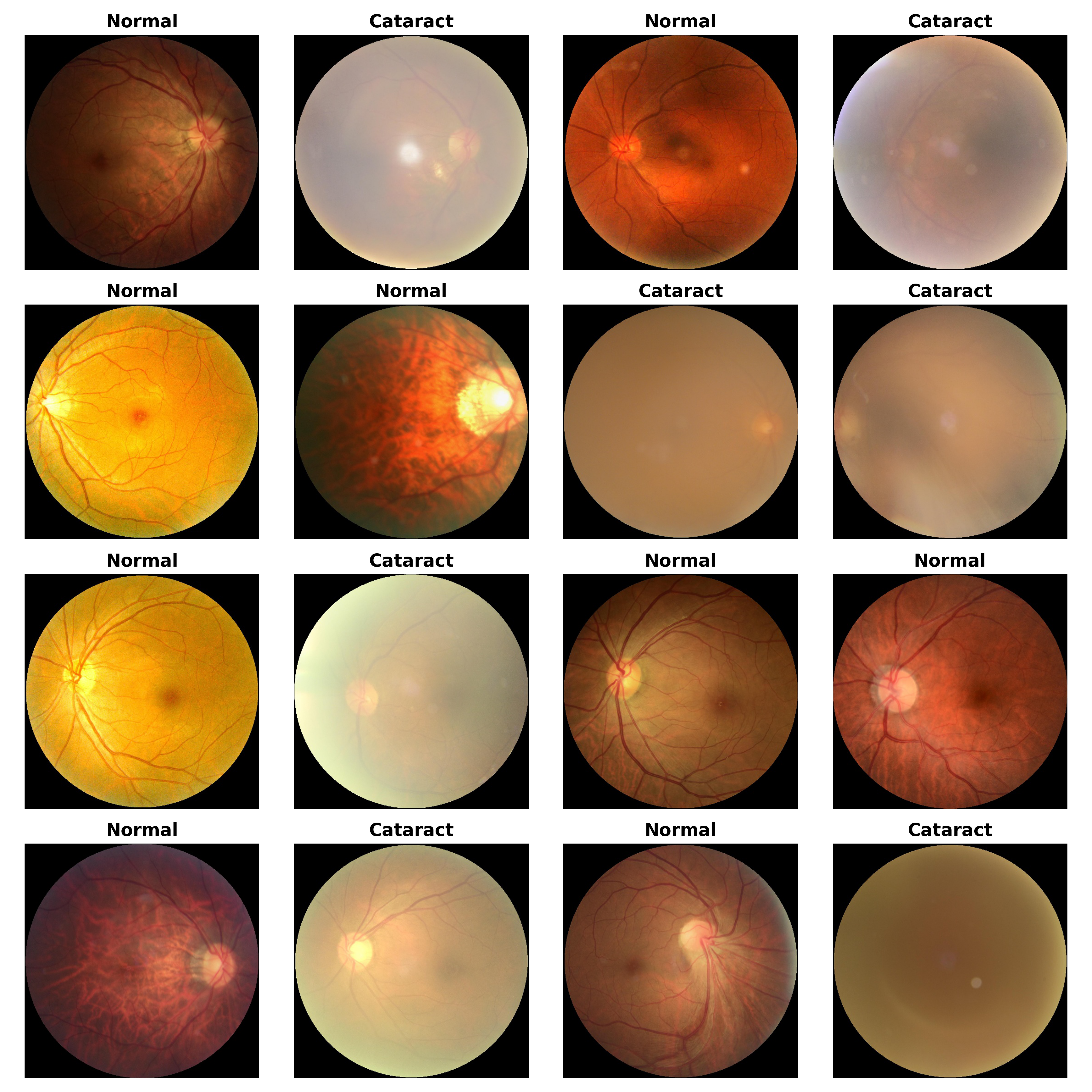}
		\caption{Fundus image samples: \texttt{normal} and \texttt{cataract} eyes}
	\end{center}
\end{figure}

\subsection{Dual-Eye Dataset}

To utilize paired fundus images per patient, we created a custom PyTorch dataset, \texttt{DualEyeFundusDataset}, which returns both left and right eye images with a single binary label. We filtered out ambiguous or noisy cases, keeping only samples where both eyes were clearly labeled as either Normal Fundus (0) or Cataract (1). At retrieval, both images are loaded and preprocessed identically. The pair’s label is set to 1 (cataract) if either eye has cataract; otherwise, it is 0 (normal). This dual-eye approach lets us explore if combining information from both eyes improves classification compared to single-eye models. With this filtering and labeling strategy, the total number of dual-eye samples available for training and evaluation was 2,254.

\subsection{Preprocessing}

To prepare the fundus images for input into the neural networks, we applied a series of preprocessing steps using \texttt{torchvision.transforms}, designed to standardize inputs and enhance model robustness. All images were resized to $224 \times 224$ pixels to ensure compatibility with standard backbone architectures such as Vision Transformers and EfficientNet. To improve generalization and reduce overfitting, we applied data augmentation techniques including random resized cropping, horizontal flipping, small-angle rotations within $\pm15^\circ$, and color jittering (adjustments to brightness and contrast). After resizing and augmentation, images were normalized using ImageNet mean and standard deviation statistics (\ref{eq:mean_std}):

\begin{equation}\label{eq:mean_std}
	\mu     = [0.485, 0.456, 0.406]  \quad
	\sigma  = [0.229, 0.224, 0.225]
\end{equation}

\quad

For label encoding, we formulated a binary classification task by selecting only samples labeled as \textit{``Normal''} or \textit{``Cataract''}, encoding these labels appropriately for training. The dataset was split into training and validation subsets using an 80/20 stratified split to preserve class balance. PyTorch \texttt{DataLoader} objects were employed to create mini-batches and shuffle data during training and evaluation phases.

\begin{figure}[hbt]
	\begin{center}
		\setlength{\unitlength}{0.0105in}%
		\includegraphics[width=0.9\linewidth]{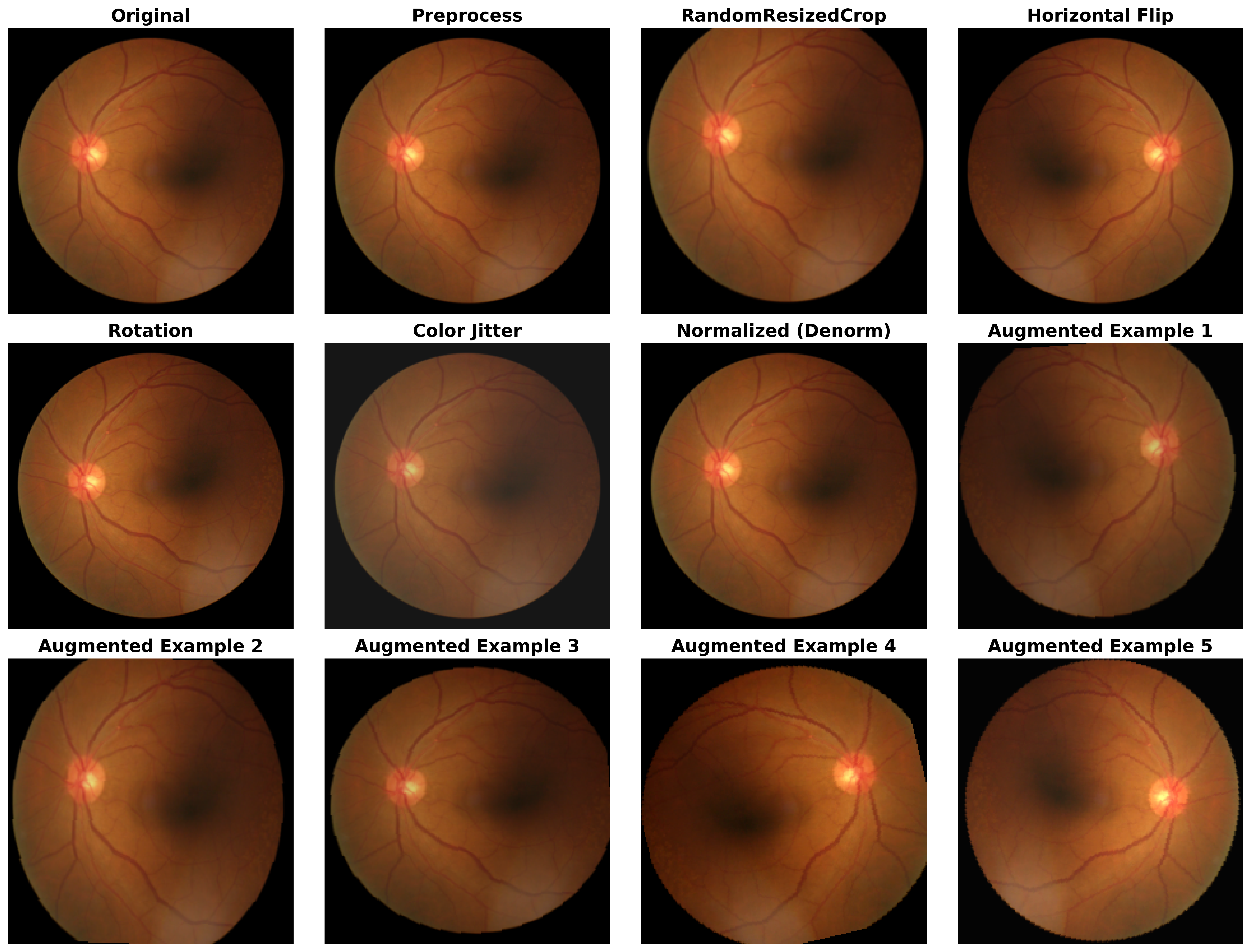}
		\caption{Visualization of fundus image preprocessing and data augmentation effects on a single sample}
	\end{center}
\end{figure}

\subsection{Models}

In this subsection, we describe the deep learning architectures, training regimes, and optimization strategies employed for cataract detection.
Our pipeline includes both single-eye and dual-eye networks, as well as a knowledge distillation framework to create lightweight yet accurate models. The complete list of models, their types, and key characteristics is summarized in Table~\ref{tab:model_summary}. All experiments were conducted under two fine-tuning regimes: (i) Fully fine-tuned — all parameters updated; (ii) Frozen backbone — all feature extractor weights frozen, training only the classification head.

\subsubsection{Convolutional Neural Networks (CNNs)}
We evaluated three widely used CNN architectures from the \texttt{timm} library, pretrained on ImageNet-1k: 
\begin{itemize}
	\item \textbf{ResNet-50} \cite{he2015deepresiduallearningimage} — employs residual connections to mitigate vanishing gradients.
	\item \textbf{DenseNet-121} \cite{huang2018denselyconnectedconvolutionalnetworks} — uses dense connectivity for efficient gradient flow.
	\item \textbf{EfficientNet-B0} \cite{tan2020efficientnetrethinkingmodelscaling} — leverages compound scaling for balanced performance and efficiency.
\end{itemize}

\subsubsection{Transformer Attention-Based Models} We investigated transformer-based architectures initialized from ImageNet-pretrained weights with input resolution $224\times224$: 
\begin{itemize}
	\item \textbf{ViT-Tiny} and \textbf{ViT-Base} \cite{dosovitskiy2020image} — utilize patch embeddings and global self-attention.
	\item \textbf{Swin-Base} \cite{liu2021swintransformerhierarchicalvision} — employs hierarchical shifted-window attention.
	\item \textbf{DeiT-Tiny} and \textbf{DeiT-Base} \cite{touvron2021trainingdataefficientimagetransformers} — optimized for smaller datasets via distillation-aware pretraining.
\end{itemize}

\subsubsection{Lightweight Models}
For resource-constrained deployment, we employed \textbf{MobileNetV2} \cite{howard2019searchingmobilenetv3} due to its low parameter count. Both full fine-tuning and frozen-backbone training were applied.

\subsubsection{Knowledge Distillation}
Knowledge distillation transferred knowledge from the best-performing Swin-Base single-eye model (teacher) into MobileNetV2 (student), using a weighted combination of soft-target Kullback–Leibler divergence and hard-label cross-entropy loss (Equation~\ref{eq:loss}).

\begin{equation}\label{eq:loss}
	\mathcal{L} = \alpha \cdot \mathrm{KL}\left(\sigma\left(\frac{z_s}{T}\right), \sigma\left(\frac{z_t}{T}\right)\right) \cdot T^2 + (1-\alpha) \cdot \mathrm{CE}(z_s, y)
\end{equation}

where $z_s$ and $z_t$ are the student and teacher logits, $T=2.0$ is the temperature, and $\alpha=0.7$ the blending factor.

\subsection{Siamese Dual-Eye Model}
\begin{figure*}[hbt]
	\begin{center}
		\setlength{\unitlength}{0.0105in}%
		\includegraphics[width=0.7\linewidth]{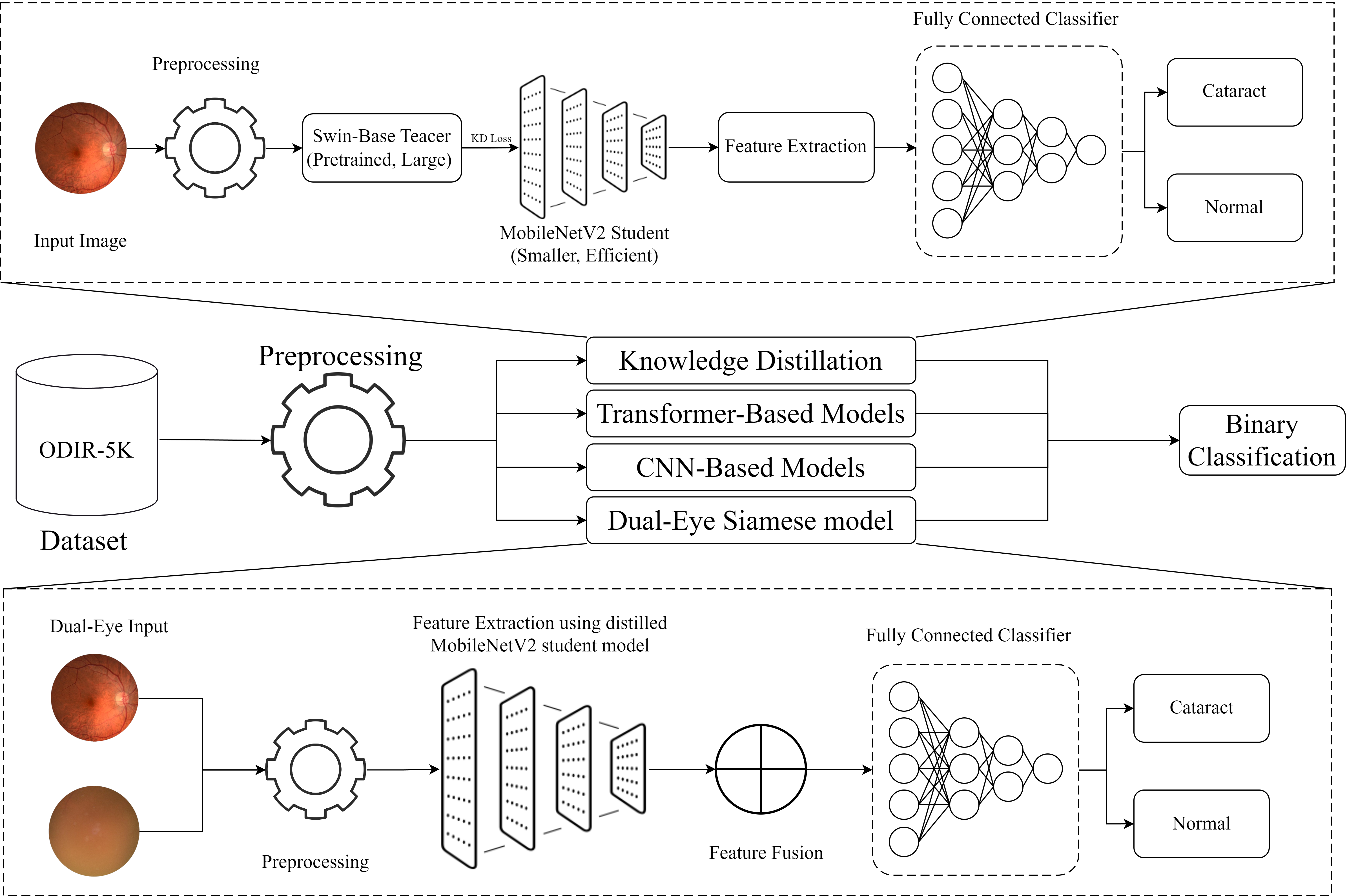}
		\caption{Overview of the Cataract Detection Framework Using Knowledge Distillation and Dual-Eye Models.
			The diagram illustrates the complete workflow starting from the ODIR-5K dataset preprocessing, followed by training multiple classification approaches including transformer-based models, CNN-based models, and a knowledge-distilled MobileNetV2 student model derived from a pretrained Swin-Base teacher. The figure also highlights the dual-eye Siamese architecture where features extracted separately from each eye using the distilled MobileNetV2 are fused and fed into a fully connected classifier for final binary classification between cataract and normal cases.}
			\label{fig:framework}
	\end{center}
\end{figure*}

Our dual-eye classification architecture utilizes a Siamese design based on the distilled MobileNetV2 backbone, aiming to leverage the complementary information inherent in paired left and right fundus images (Figure~\ref{fig:framework}). The model accepts two separate images—one from each eye—and processes them through the same feature extractor with shared weights, producing 1280-dimensional feature vectors for each eye. These are projected to 128 dimensions via a fully connected layer with ReLU activation and dropout, concatenated into a 256-dimensional vector, and passed through a two-layer MLP classifier (hidden layer of 64 units, dropout) to output the final binary prediction. Following the clinical assumption that cataract in either eye warrants a positive classification, the model fuses inter-eye features to capture symmetry and complementary cues, leading to improved performance compared to single-eye models.

\subsection{Training Details}

Models were trained using the Cross-Entropy loss function, which is well-suited for binary classification tasks. Optimization was performed with the AdamW optimizer, employing a weight decay of $1 \times 10^{-5}$ to promote regularization and stabilize training. To refine the learning process, we utilized a ReduceLROnPlateau scheduler that monitored validation accuracy and reduced the learning rate by a factor of 0.5 if no improvement was observed over two consecutive epochs. Additionally, early stopping was applied to prevent overfitting. Training was halted if validation accuracy failed to improve for five consecutive epochs, ensuring retention of the best-performing model checkpoint.

\begin{table}[htbp]
	\begin{center}
		\caption{Training hyperparameters}
		\label{tab:hyperparams}
		{\tt
			\begin{tabular}{ll}
				\toprule
				\textbf{Setting} & \textbf{Value} \\
				\midrule
				Input resolution & $224\times224$ \\
				Batch size       & 16 \\
				Optimizer        & AdamW \\
				Learning rate    & $1 \times 10^{-4}$ \\
				Weight decay     & $1 \times 10^{-5}$ \\
				Label smoothing  & $0.1$ \\
				Scheduler        & ReduceLROnPlateau (factor of 0.5) \\
				Early stopping   & Patience 5 \\
				Training regime  & Full fine-tune / Full frozen \\
				\bottomrule
			\end{tabular}
		}
	\end{center}
\end{table}

\subsection{Explainability}

To improve interpretability of our cataract detection models, we applied Gradient-weighted Class Activation Mapping (Grad-CAM) on CNN architectures. Grad-CAM produces heatmaps that highlight regions in fundus images most influential to the model’s predictions. Using the \texttt{pytorch-grad-cam} library, we targeted the last convolutional layer of each CNN (e.g., \texttt{features[-1]} in DenseNet-121). Sample Grad-CAM visualizations for cataract and normal fundus cases are provided in Appendix A, figures \ref{gradcamcataract} and \ref{gradcamnormal}, respectively.

\begin{table*}[hbt]
	\small
	\centering
	\caption{Summary of deep learning models used for cataract detection.}
	\label{tab:model_summary}
	\begin{tabular}{lll c c}
		\toprule
		\textbf{Model} & \textbf{Type} & \textbf{Pretrained} & \textbf{Key Features} \\
		\midrule
		ResNet-50          & CNN           & ImageNet-1k       & Residual connections \\
		DenseNet-121       & CNN           & ImageNet-1k       & Dense connectivity \\
		EfficientNet-B0    & CNN           & ImageNet-1k       & Compound scaling \\
		ViT-Tiny / ViT-Base & Transformer   & ImageNet-1k       & Patch embeddings, global attention \\
		Swin-Base          & Transformer   & ImageNet-1k       & Shifted-window attention \\
		DeiT-Tiny / DeiT-Base & Transformer & ImageNet-1k       & Distillation-aware pretraining \\
		MobileNetV2        & Lightweight CNN & ImageNet-1k     & Low parameter count \\
		Distilled MobileNetV2 & Lightweight CNN & From Swin-Base & Knowledge distillation from Swin \\
		Dual-eye Siamese (MobileNetV2) & Lightweight CNN & From distilled MobileNetV2 & Shared backbone, paired input \\
		\bottomrule
	\end{tabular}
\end{table*}

\qquad

\section{Experiments and Results}

We conducted a thorough evaluation of multiple deep learning models for cataract classification using the curated fundus dataset. The models were assessed using key metrics including accuracy and F1-score, to effectively handle the class imbalance present in the data. Among all tested architectures, the Swin-Base transformer model achieved the highest overall performance, reaching an accuracy of 98.58\% and an F1-score of 0.9857. This strong result demonstrates the power of hierarchical self-attention mechanisms in capturing relevant retinal features for cataract detection (bolded see Table \ref{tab:model_performance_slim}).

To address computational efficiency alongside accuracy, we employed knowledge distillation to transfer learned representations from the Swin-Base teacher model—which has 87M parameters—to a lightweight MobileNetV2 student model with only 2,562 parameters when fully frozen. Remarkably, the distilled MobileNetV2 maintained competitive performance with 98.42\% accuracy and an F1-score of 0.9787, while substantially reducing model size and inference cost, making it highly suitable for resource-limited clinical environments (see Table \ref{tab:model_performance_slim}).

Further, we introduced a novel dual-eye Siamese network built upon the distilled MobileNetV2 backbone. By simultaneously leveraging paired left and right fundus images, this model achieved an accuracy of 98.21\%, confirming the benefit of integrating bilateral ocular information for improved diagnostic accuracy. Table \ref{tab:model_performance_slim} provides a detailed summary of the models’ performances, training regimes (full fine-tuning versus frozen backbone), F1 scores, and parameter counts. It is evident that the knowledge distillation approach offers an excellent balance between accuracy and model compactness. The dual-eye Siamese network, while slightly lower in F1, demonstrates a promising avenue for leveraging multi-view data in ophthalmic AI applications.

A comparative analysis of the two training regimes across all architectures is provided in Appendix~E (Fig.~\ref{ablation}), highlighting the consistent advantage of full fine-tuning in both accuracy and F1-score. The training dynamics of these models are illustrated in Appendix~C (Fig.~\ref{curves}), showing convergence trends and validation behavior. Additionally, Appendix~D (Fig.~\ref{ROC}) presents ROC curves comparing classification performance across all top-performing models, providing a detailed view of sensitivity and specificity trade-offs.

\begin{sidewaystable}[htbp]
	\begin{center}
		\setlength{\tabcolsep}{8pt}
		\caption{\centering Model performance (Accuracy, F1 Score) and parameter counts under different training regimes. The first value in Accuracy and F1 Score columns corresponds to full fine-tuning (FT) and the second to full frozen (FR).}
		\label{tab:model_performance_slim}
		\begin{tabular}{@{}llccc ccc@{}}
			\toprule
			\textbf{Category} & \textbf{Model Name} & \textbf{Accuracy (FT / FR)} & \textbf{F1 Score (FT / FR)} & \textbf{Params FT} & \textbf{Params FR} \\
			\midrule
			\multirow{3}{*}{CNN-based}
			& ResNet-50           & 0.9763 / 0.9069 & 0.9782 / 0.8627 & 23,512,130  & 4,098   \\
			& DenseNet-121        & 0.9795 / 0.9653 & 0.9824 / 0.9672 & 6,955,906   & 2,050   \\
			& EfficientNet-B0     & 0.9795 / 0.9590 & 0.9740 / 0.9706 & 4,010,110   & 2,562   \\
			\midrule
			\multirow{5}{*}{Transformer-based}
			& ViT-Tiny            & 0.9842 / 0.9527 & 0.9808 / 0.9576 & 5,524,802   & 386     \\
			& ViT-Base            & 0.9748 / 0.9716 & 0.9709 / 0.9774 & 85,800,194  & 1,538   \\
			& \textbf{Swin-Base}           & \textbf{0.9858} / 0.9842 & \textbf{0.9857} / 0.9836 & 86,745,274  & 2,050   \\
			& DeiT-Base           & 0.9763 / 0.9795 & 0.9755 / 0.9804 & 85,800,194  & 1,538   \\
			& DeiT-Tiny           & 0.9732 / 0.9606 & 0.9742 / 0.9598 & 5,524,802   & 386     \\
			\midrule
			Lightweight CNN & MobileNetV2         & 0.9795 / 0.9669 & 0.9823 / 0.9751 & 2,226,434   & 2,562   \\
			Knowledge Distillation & Swin + MobileNet & 0.9874 / 0.9842 & 0.9787 / 0.9754 & 2,226,434   & 2,562   \\
			Dual-Eye        & Siamese + MobileNet  &  --- / 0.9821    & --- / 0.8601    & 2,404,418   & ---     \\
			\bottomrule
		\end{tabular}
	\end{center}
\end{sidewaystable}

\section{Discussion}

Detecting cataracts from fundus images is relatively straightforward because of clear visual signs like lens opacity and central blur. Compared to other retinal diseases, such as diabetic retinopathy, which often involve subtle changes, these features make classification easier. This visual clarity likely helped the models achieve strong overall performance.

\subsection{Model Efficiency and Scalability}
Transformer-based architectures, especially the Swin-Base model ($\sim$86M parameters), showed impressive accuracy and robustness. However, their high computational and memory requirements make them challenging to deploy in real-world clinical settings, particularly in resource-limited environments. On the other hand, the frozen MobileNetV2, used as the student in knowledge distillation ($\sim$2.5M parameters), strikes a strong balance between performance and efficiency. Despite having far fewer parameters, this lightweight model delivers competitive accuracy and F1-scores, making it a practical choice for scalable cataract screening.

\subsection{Dual-Eye Siamese Model}
Our novel dual-eye Siamese model takes advantage of the natural symmetry and complementary information between both eyes. By processing paired fundus images together, it achieved better classification results than some single-eye models, underscoring the clinical value of integrating data from both eyes. Importantly, this dual-eye system remains lightweight, as it uses the knowledge-distilled MobileNetV2 backbone obtained from our KD step, making it practical for real-world deployment.

\subsection{Explainability and Grad-CAM Insights}
Grad-CAM analyses on our CNN models (Figures \ref{gradcamcataract} and \ref{gradcamnormal}) showed that the networks consistently focused on clinically relevant areas. For cataract cases, the attention concentrated on regions with lens opacity and central blur, while for normal fundus images, the main activations overlapped with regions dense with blood vessels, reflecting the natural anatomical structures. This visual confirmation not only builds confidence in the model’s decisions but also supports potential clinical adoption by providing interpretable evidence aligned with expert knowledge.

\quad

To our knowledge, this study is among the first to integrate knowledge distillation with dual-eye modeling specifically for cataract detection using fundus images. Beyond proposing this novel architecture, we performed extensive benchmarking, training over 10 different model types and experimenting with more than 20 training regimes, to ensure robust evaluation. This combined approach not only advances methodological innovation but also addresses practical considerations of efficiency and interpretability, paving the way for more accessible and reliable automated screening tools.

\section{Conclusion and Future Work}

In this study, we developed a comprehensive deep learning pipeline for automated cataract classification using retinal fundus images. Our experiments demonstrated that advanced models such as the Swin-Base transformer and a knowledge-distilled MobileNetV2 can achieve high accuracy while maintaining model interpretability through explainability techniques. The introduction of a novel Siamese dual-eye architecture further improved performance by effectively leveraging paired left and right eye inputs, reflecting the clinical practice of bilateral assessment. Explainability analyses provided valuable insights into the models’ decision-making processes, confirming that predictions are grounded in clinically relevant features such as lens opacity and central blur. This transparency supports trust and potential integration into clinical workflows. Looking ahead, future research will focus on extending this framework to multi-task classification encompassing various ocular diseases, enabling broader screening capabilities from fundus images. To further enhance model robustness and address data scarcity, advanced data augmentation techniques and synthetic data generation using generative models will be explored. Additionally, incorporating multimodal approaches—combining fundus images with other clinical data such as patient demographics, OCT scans, or genetic information—holds promise for improving diagnostic accuracy and providing a more comprehensive understanding of ocular health.

\section*{Appendix}

\subsection{Appendix – comparative overview of previous studies}
This appendix presents a comparative overview of previous studies on cataract detection and related ophthalmic disease classification using deep learning. Table \ref{tab:related_work} summarizes the datasets, imaging modalities, model architectures, and reported performance metrics for each study, highlighting the diversity of approaches and benchmark results in the literature.

\begin{sidewaystable}[b]
	\small
	\centering
	\caption{\centering Representative related works in ophthalmic AI. A dash (—) indicates metric not reported here or varies across datasets.}
	\label{tab:related_work}
	\setlength{\tabcolsep}{6pt}
	\begin{tabular}{@{}p{3cm}llllp{5cm}@{}}
		\toprule
		\textbf{Study} & \textbf{Disease / Task} & \textbf{Modality} & \textbf{Approach \& Metric} & \textbf{Notes} \\
		\midrule
		\cite{Atwany2022-qs} & DR screening (survey) & Fundus & Survey / taxonomy, — & Comprehensive review of DL trends and datasets for DR. \\
		\cite{Tsiknakis2021-oc} & DR grading & Fundus & CNN variants, Expert-level & Reports expert-level performance for DR diagnosis/grading. \\
		\cite{Alyoubi2020-mz} & DR detection & Fundus & CNN, High accuracy & Strong screening performance on fundus images. \\
		\cite{Deng2022-bx} & AMD detection & OCT / Fundus & DL classifiers, High sensitivity/specificity & Sometimes surpasses human experts. \\
		\cite{Koseoglu2023-ms} & Glaucoma detection & OCT / Fundus & DL models, — & High diagnostic performance reported. \\
		\cite{Sheng2022-jd} & Glaucoma / AMD & OCT / Fundus & DL models, — & Competitive with or above human experts in some settings. \\
		\cite{Gonz_lez_Gonzalo_2019} & Multi-disease (DR+AMD) & Fundus & Multi-task CNN, — & Integrated multi-disease screening framework. \\
		\cite{zhang2022machinelearningcataractclassification} & Cataract (survey) & Fundus / Slit-lamp & Survey, — & Highlights limited fundus-based cataract DL vs. other modalities. \\
		\cite{Elloumi2021-gm} & Cataract detection & Fundus & CNN (MobileNetV2), $>$90\% acc. & Lightweight model suitable for deployment. \\
		\cite{Xie2023-ha} & Cataract detection & Fundus & CNN (DenseNet), $>$90\% acc. & Shows strong cataract classification with DenseNet. \\
		\cite{padalia2022cnnlstmcombinationnetworkcataract} & Cataract detection & Fundus & CNN+LSTM hybrid, $>$90\% acc. & Combines convolutional features with sequence-style head. \\
		\cite{dosovitskiy2021imageworth16x16words} & Generic vision & — & ViT (Transformer), — & Patch-based global attention; inspires medical imaging use. \\
		\cite{liu2021swintransformerhierarchicalvision} & Generic vision & — & Swin Transformer, — & Hierarchical shifted-window attention. \\
		\cite{touvron2021trainingdataefficientimagetransformers} & Generic vision & — & DeiT (distillation), — & Data-efficient training via distillation. \\
		\cite{Selvaraju_2019} & Explainability & — & Grad-CAM, — & Localizes decision-relevant regions for clinical trust. \\
		\bottomrule
	\end{tabular}
\end{sidewaystable}

\clearpage
\onecolumn

{\centering \subsection{Appendix – Grad-CAM}
	
{\centering This appendix presents Gradient-weighted Class Activation Mapping (Grad-CAM) visualizations for the key convolutional neural network models evaluated in this study. The heatmaps highlight regions within the fundus images that contributed most to the models’ cataract classification decisions. These visual explanations provide insight into the models’ interpretability and alignment with clinically relevant features such as lens opacity and blur areas. }

\begin{figure}[hbt]
	\begin{center}
		\setlength{\unitlength}{0.0105in}%
		\includegraphics[width=0.8\linewidth]{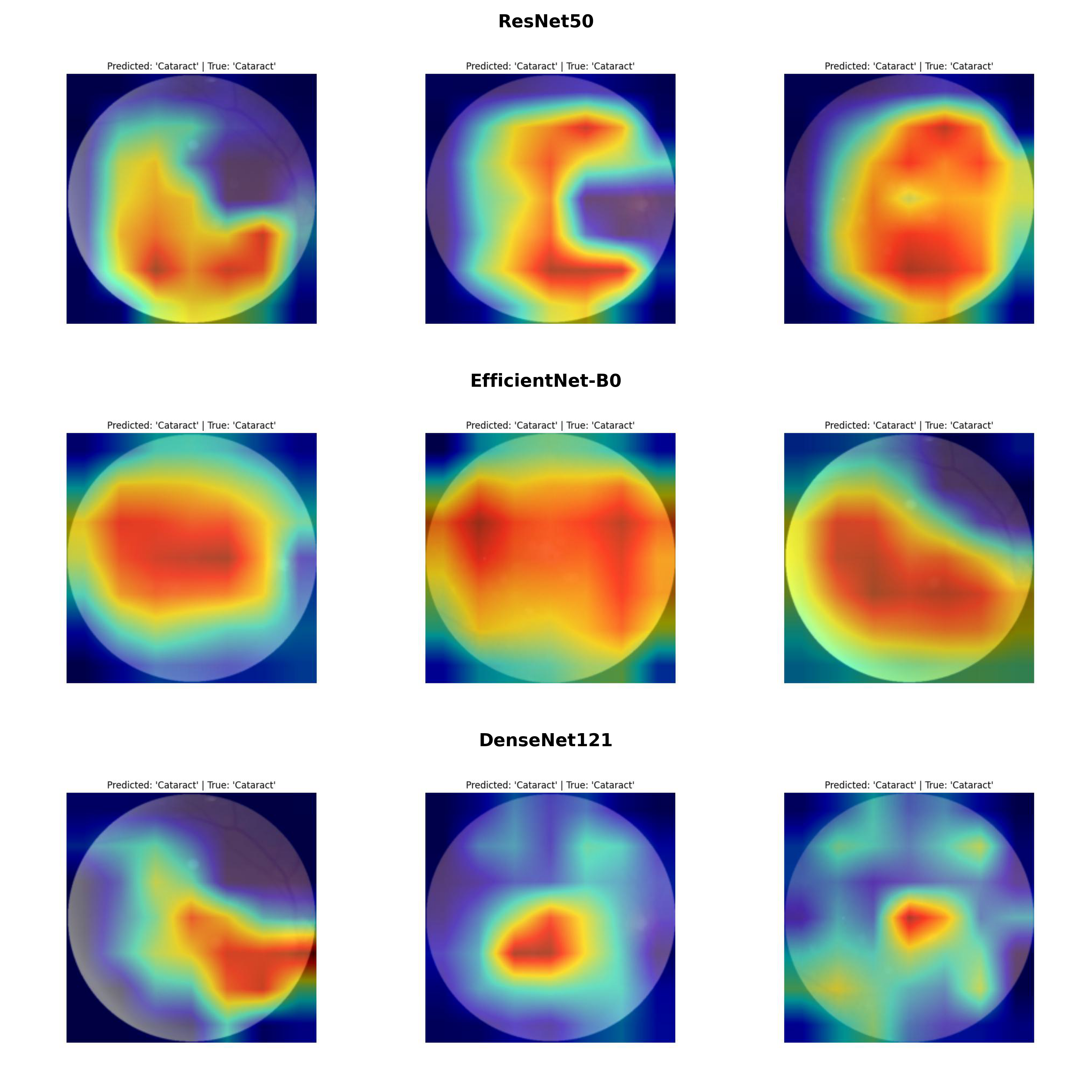}
		\caption{Grad-CAM visualizations of CNN models on cataract fundus images: ResNet50, EfficientNet-B0, and DenseNet121}
		\label{gradcamcataract}
	\end{center}
\end{figure}

\begin{figure}[hbt]
	\begin{center}
		\setlength{\unitlength}{0.0105in}%
		\includegraphics[width=0.8\linewidth]{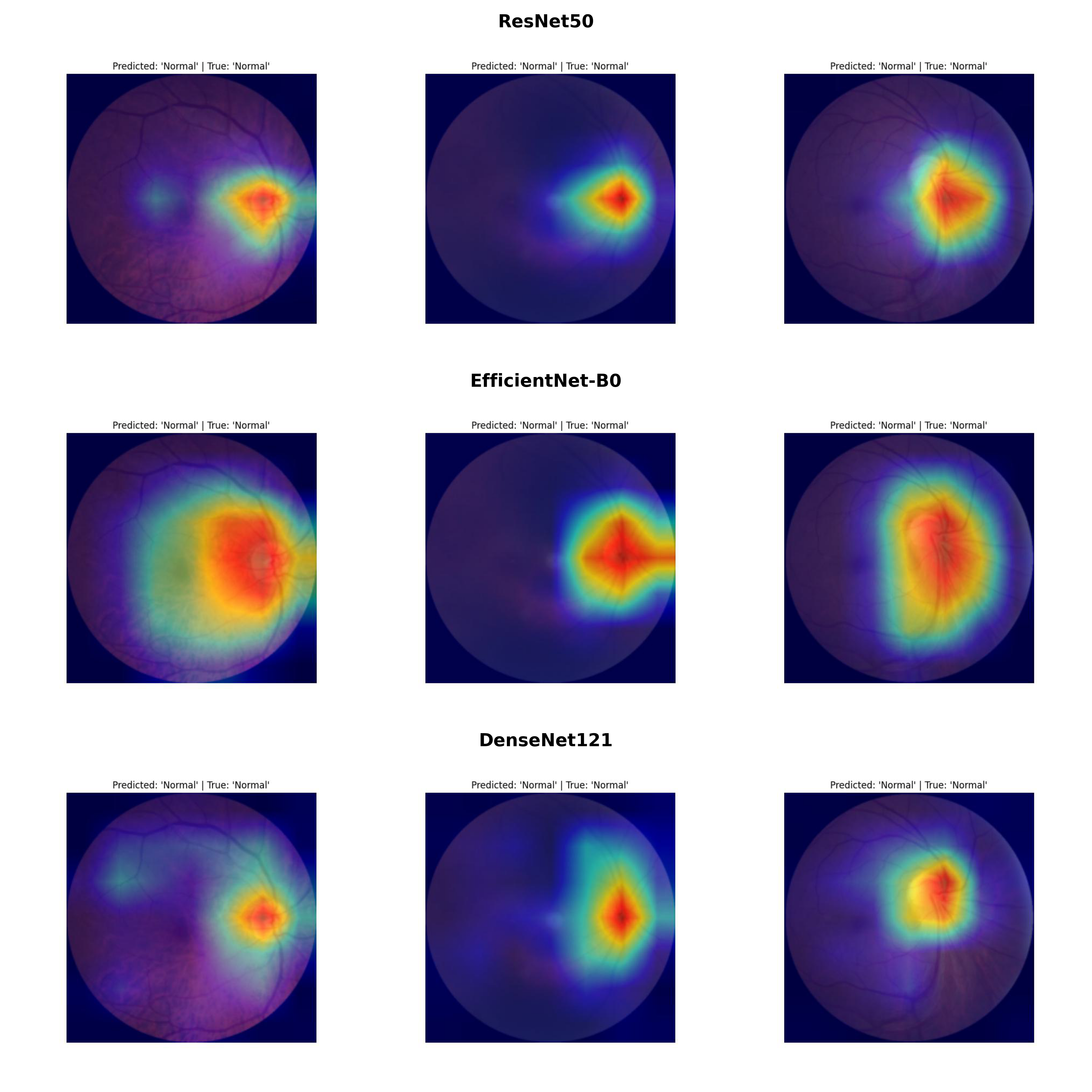}
		\caption{Grad-CAM visualizations of CNN models on normal fundus images: ResNet50, EfficientNet-B0, and DenseNet121}
		\label{gradcamnormal}
	\end{center}
\end{figure}

\clearpage

{\centering \subsection{Appendix – Learning Curves for Top Models}

{\centering
	This appendix shows the Learning curves for the best-performing models.\par
}

\begin{figure*}[hbt]
	\begin{center}
		\setlength{\unitlength}{0.0105in}%
		\includegraphics[width=0.8\linewidth]{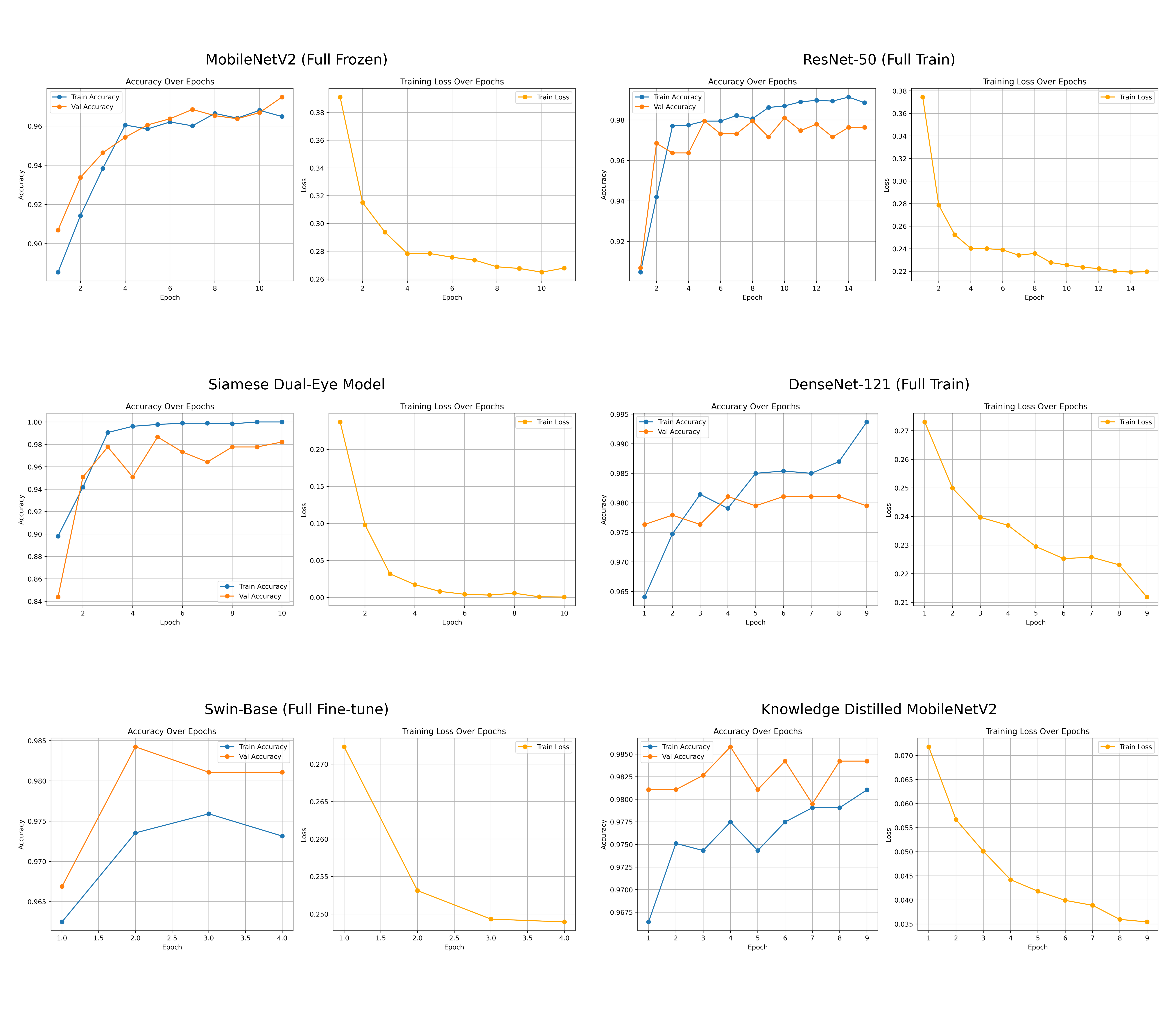}
		\captionsetup{justification=centering}
		\caption{Learning curves of the top-performing deep learning models for cataract detection.}
		\label{curves}
	\end{center}
\end{figure*}

\clearpage

{\centering \subsection{Appendix – ROC Curves of Top Models}

{\centering
	This appendix shows the ROC curves for the best-performing models.\par
}

\begin{figure*}[hbt]
	\begin{center}
		\setlength{\unitlength}{0.0105in}%
		\includegraphics[width=0.8\linewidth]{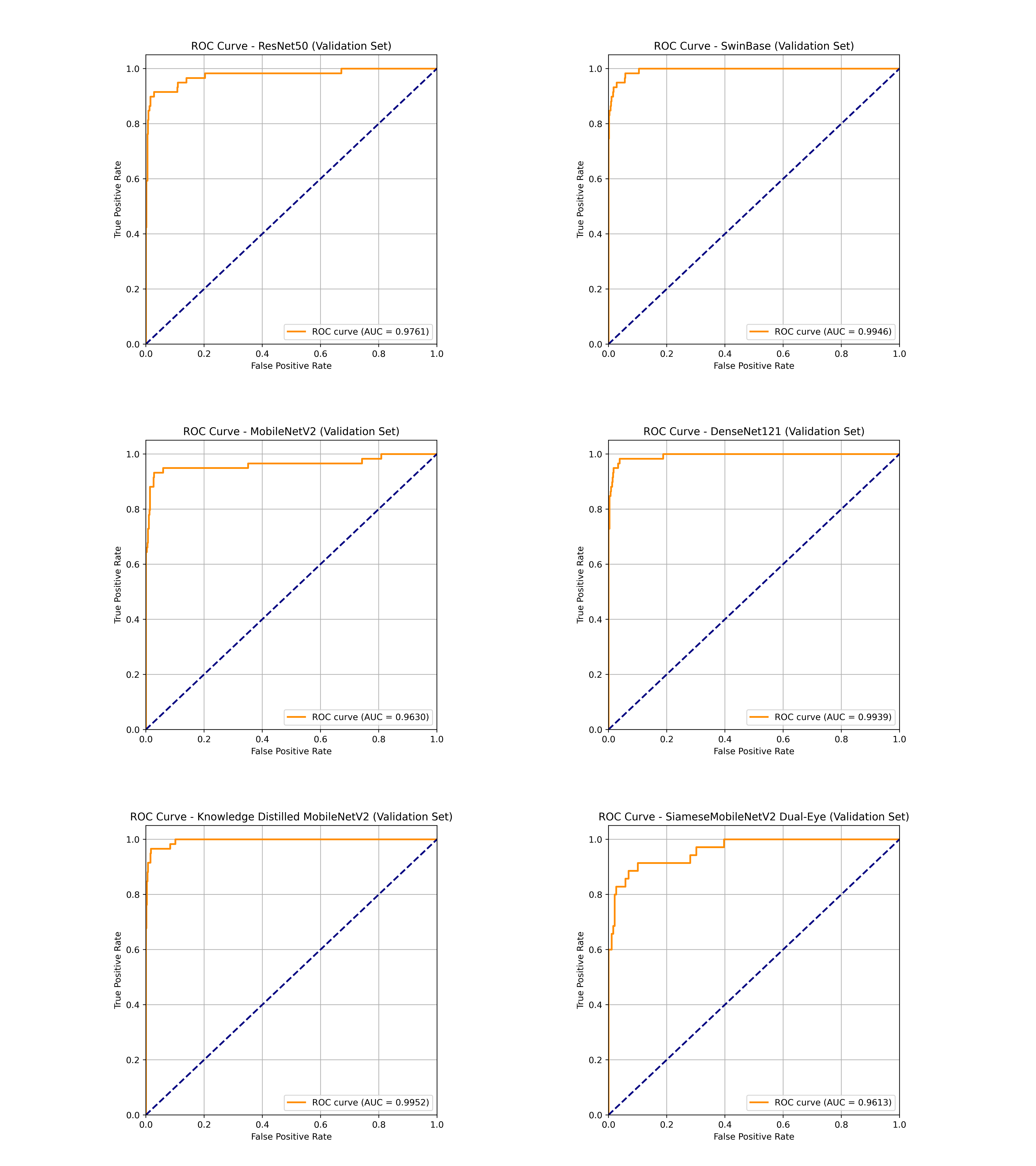}
		\captionsetup{justification=centering}
		\caption{ROC curves of the top-performing deep learning models for cataract detection. Each subplot shows the classification performance measured across different thresholds, highlighting the comparative strengths of CNNs, transformers, knowledge distillation, and dual-eye architectures.}
		\label{ROC}
	\end{center}
\end{figure*}

\clearpage

{\centering \subsection{Appendix – Ablation Study: Training Regimes Performance Comparison}

{\centering This appendix presents a detailed comparison of model performance under two training regimes: full fine-tuning and full frozen backbone training. The figure shows accuracy and F1-score metrics across a range of deep learning architectures, including CNNs, transformers, lightweight models, and knowledge-distilled variants. These results highlight the trade-offs between training complexity and predictive effectiveness, emphasizing how knowledge distillation and dual-eye models perform relative to standard approaches.}

\begin{figure*}[hbt]
	\begin{center}
		\setlength{\unitlength}{0.0105in}%
		\includegraphics[width=0.9\linewidth]{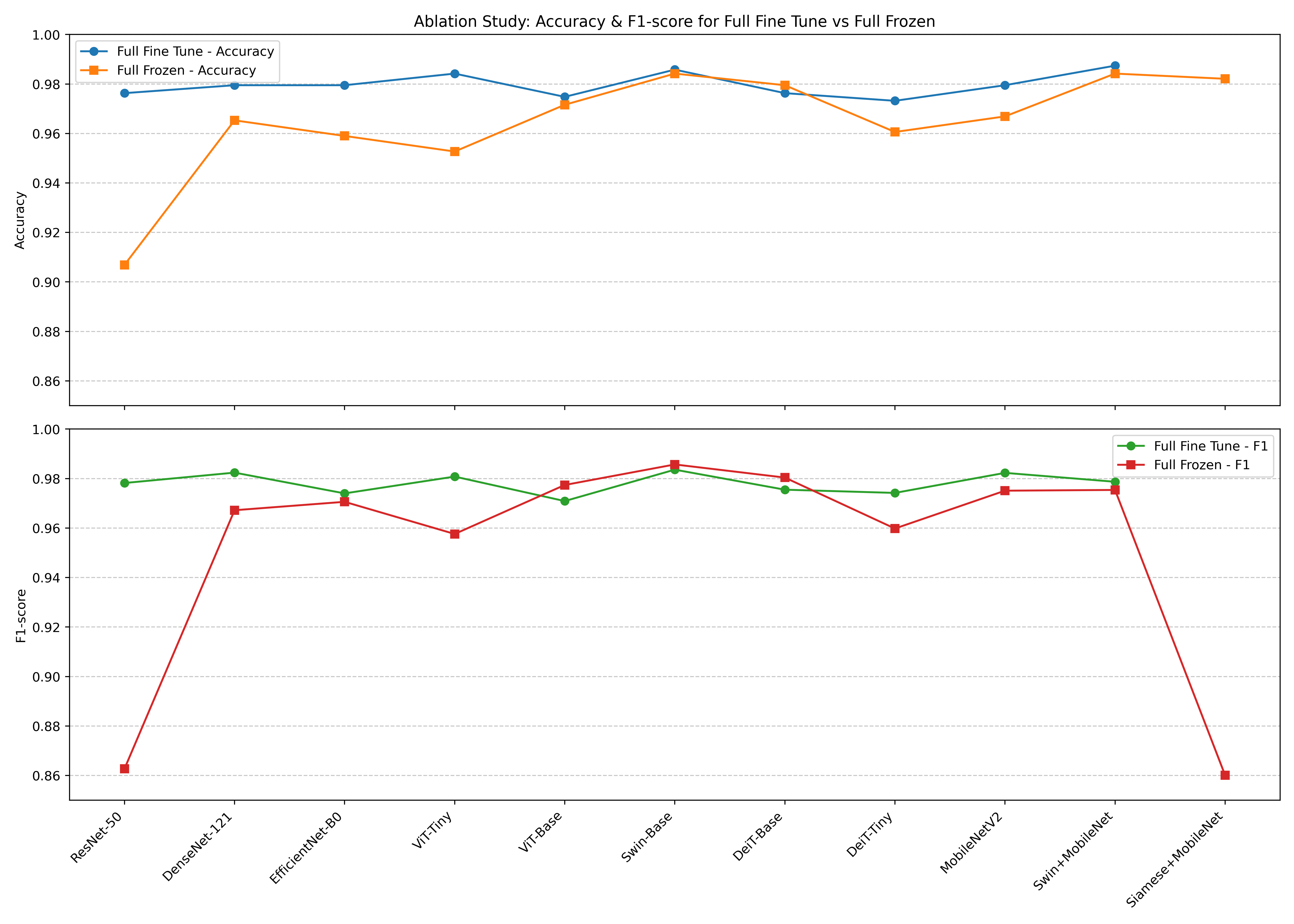}
		\captionsetup{justification=centering}
		\caption{Performance Comparison of Full Fine-Tuning vs Frozen Backbone Across Models}
		\label{ablation}
	\end{center}
\end{figure*}

\twocolumn

\bibliographystyle{plainnat}
\bibliography{references}

\begin{thebibliography}{37}
\providecommand{\natexlab}[1]{#1}
\providecommand{\url}[1]{\texttt{#1}}
\expandafter\ifx\csname urlstyle\endcsname\relax
  \providecommand{\doi}[1]{doi: #1}\else
  \providecommand{\doi}{doi: \begingroup \urlstyle{rm}\Url}\fi

\bibitem[Alyoubi et~al.(2020)Alyoubi, Shalash, and Abulkhair]{Alyoubi2020-mz}
Wejdan~L Alyoubi, Wafaa~M Shalash, and Maysoon~F Abulkhair.
\newblock Diabetic retinopathy detection through deep learning techniques: A
  review.
\newblock \emph{Inform. Med. Unlocked}, 20\penalty0 (100377):\penalty0 100377,
  2020.

\bibitem[Atwany et~al.(2022)Atwany, Sahyoun, and Yaqub]{Atwany2022-qs}
Mohammad~Z Atwany, Abdulwahab~H Sahyoun, and Mohammad Yaqub.
\newblock Deep learning techniques for diabetic retinopathy classification: A
  survey.
\newblock \emph{IEEE Access}, 10:\penalty0 28642--28655, 2022.

\bibitem[Bilal et~al.(2025)Bilal, Keles, and Bendechache]{Bilal2025-fw}
Hazrat Bilal, Ayse Keles, and Malika Bendechache.
\newblock Advances in disease detection through retinal imaging: A systematic
  review.
\newblock \emph{Comput. Biol. Med.}, 194\penalty0 (110412):\penalty0 110412,
  August 2025.

\bibitem[Chua et~al.(2017)Chua, Lim, Fenwick, Gan, Tan, Lamoureux, Mitchell,
  Wang, Wong, and Cheng]{Chua2017-ui}
Jacqueline Chua, Blanche Lim, Eva~K Fenwick, Alfred Tau~Liang Gan, Ava~Grace
  Tan, Ecosse Lamoureux, Paul Mitchell, Jie~Jin Wang, Tien~Yin Wong, and
  Ching-Yu Cheng.
\newblock Prevalence, risk factors, and impact of undiagnosed visually
  significant cataract: The singapore epidemiology of eye diseases study.
\newblock \emph{PLoS One}, 12\penalty0 (1):\penalty0 e0170804, January 2017.

\bibitem[Deng et~al.(2022)Deng, Qiao, Du, Qu, Wan, Li, and Huang]{Deng2022-bx}
Yanhui Deng, Lifeng Qiao, Mingyan Du, Chao Qu, Ling Wan, Jie Li, and Lulin
  Huang.
\newblock Age-related macular degeneration: Epidemiology, genetics,
  pathophysiology, diagnosis, and targeted therapy.
\newblock \emph{Genes Dis.}, 9\penalty0 (1):\penalty0 62--79, January 2022.

\bibitem[Dosovitskiy et~al.(2021{\natexlab{a}})Dosovitskiy, Beyer, Kolesnikov,
  Weissenborn, Zhai, Unterthiner, Dehghani, Minderer, Heigold, Gelly,
  Uszkoreit, and Houlsby]{dosovitskiy2020image}
Alexey Dosovitskiy, Lucas Beyer, Alexander Kolesnikov, Dirk Weissenborn,
  Xiaohua Zhai, Thomas Unterthiner, Mostafa Dehghani, Matthias Minderer, Georg
  Heigold, Sylvain Gelly, Jakob Uszkoreit, and Neil Houlsby.
\newblock An image is worth 16x16 words: Transformers for image recognition at
  scale.
\newblock In \emph{International Conference on Learning Representations
  (ICLR)}, 2021{\natexlab{a}}.

\bibitem[Dosovitskiy et~al.(2021{\natexlab{b}})Dosovitskiy, Beyer, Kolesnikov,
  Weissenborn, Zhai, Unterthiner, Dehghani, Minderer, Heigold, Gelly,
  Uszkoreit, and Houlsby]{dosovitskiy2021imageworth16x16words}
Alexey Dosovitskiy, Lucas Beyer, Alexander Kolesnikov, Dirk Weissenborn,
  Xiaohua Zhai, Thomas Unterthiner, Mostafa Dehghani, Matthias Minderer, Georg
  Heigold, Sylvain Gelly, Jakob Uszkoreit, and Neil Houlsby.
\newblock An image is worth 16x16 words: Transformers for image recognition at
  scale, 2021{\natexlab{b}}.
\newblock URL \url{https://arxiv.org/abs/2010.11929}.

\bibitem[Elloumi(2021)]{Elloumi2021-gm}
Yaroub Elloumi.
\newblock Mobile aided system of deep-learning based cataract grading from
  fundus images.
\newblock In \emph{Artificial Intelligence in Medicine}, Lecture notes in
  computer science, pages 355--360. Springer International Publishing, Cham,
  2021.

\bibitem[González‐Gonzalo et~al.(2019)González‐Gonzalo,
  Sánchez‐Gutiérrez, Hernández‐Martínez, Contreras, Lechanteur,
  Domanian, van Ginneken, and Sánchez]{Gonz_lez_Gonzalo_2019}
Cristina González‐Gonzalo, Verónica Sánchez‐Gutiérrez, Paula
  Hernández‐Martínez, Inés Contreras, Yara~T. Lechanteur, Artin Domanian,
  Bram van Ginneken, and Clara~I. Sánchez.
\newblock Evaluation of a deep learning system for the joint automated
  detection of diabetic retinopathy and age‐related macular degeneration.
\newblock \emph{Acta Ophthalmologica}, 98\penalty0 (4):\penalty0 368–377,
  November 2019.
\newblock ISSN 1755-3768.
\newblock \doi{10.1111/aos.14306}.
\newblock URL \url{http://dx.doi.org/10.1111/aos.14306}.

\bibitem[Gulshan et~al.(2016)Gulshan, Peng, Coram, Stumpe, Wu, Narayanaswamy,
  Venugopalan, Widner, Madams, Cuadros, Kim, Raman, Nelson, Mega, and
  Webster]{Gulshan2016-ru}
Varun Gulshan, Lily Peng, Marc Coram, Martin~C Stumpe, Derek Wu, Arunachalam
  Narayanaswamy, Subhashini Venugopalan, Kasumi Widner, Tom Madams, Jorge
  Cuadros, Ramasamy Kim, Rajiv Raman, Philip~C Nelson, Jessica~L Mega, and
  Dale~R Webster.
\newblock Development and validation of a deep learning algorithm for detection
  of diabetic retinopathy in retinal fundus photographs.
\newblock \emph{JAMA}, 316\penalty0 (22):\penalty0 2402, December 2016.

\bibitem[He et~al.(2015)He, Zhang, Ren, and
  Sun]{he2015deepresiduallearningimage}
Kaiming He, Xiangyu Zhang, Shaoqing Ren, and Jian Sun.
\newblock Deep residual learning for image recognition, 2015.
\newblock URL \url{https://arxiv.org/abs/1512.03385}.

\bibitem[Hinton et~al.(2015)Hinton, Vinyals, and
  Dean]{hinton2015distillingknowledgeneuralnetwork}
Geoffrey Hinton, Oriol Vinyals, and Jeff Dean.
\newblock Distilling the knowledge in a neural network, 2015.
\newblock URL \url{https://arxiv.org/abs/1503.02531}.

\bibitem[Howard et~al.(2019)Howard, Sandler, Chu, Chen, Chen, Tan, Wang, Zhu,
  Pang, Vasudevan, Le, and Adam]{howard2019searchingmobilenetv3}
Andrew Howard, Mark Sandler, Grace Chu, Liang-Chieh Chen, Bo~Chen, Mingxing
  Tan, Weijun Wang, Yukun Zhu, Ruoming Pang, Vijay Vasudevan, Quoc~V. Le, and
  Hartwig Adam.
\newblock Searching for mobilenetv3, 2019.
\newblock URL \url{https://arxiv.org/abs/1905.02244}.

\bibitem[Howard et~al.(2017)Howard, Zhu, Chen, Kalenichenko, Wang, Weyand,
  Andreetto, and Adam]{howard2017mobilenetsefficientconvolutionalneural}
Andrew~G. Howard, Menglong Zhu, Bo~Chen, Dmitry Kalenichenko, Weijun Wang,
  Tobias Weyand, Marco Andreetto, and Hartwig Adam.
\newblock Mobilenets: Efficient convolutional neural networks for mobile vision
  applications, 2017.
\newblock URL \url{https://arxiv.org/abs/1704.04861}.

\bibitem[Huang et~al.(2018)Huang, Liu, van~der Maaten, and
  Weinberger]{huang2018denselyconnectedconvolutionalnetworks}
Gao Huang, Zhuang Liu, Laurens van~der Maaten, and Kilian~Q. Weinberger.
\newblock Densely connected convolutional networks, 2018.
\newblock URL \url{https://arxiv.org/abs/1608.06993}.

\bibitem[Jin and Ye(2022)]{Jin2022-vd}
Kai Jin and Juan Ye.
\newblock Artificial intelligence and deep learning in ophthalmology: Current
  status and future perspectives.
\newblock \emph{Adv. Ophthalmol. Pract. Res.}, 2\penalty0 (3):\penalty0 100078,
  November 2022.

\bibitem[Kermany et~al.(2018)Kermany, Goldbaum, Cai, Valentim, Liang, Baxter,
  McKeown, Yang, Wu, Yan, Dong, Prasadha, Pei, Ting, Zhu, Li, Hewett, Dong,
  Ziyar, Shi, Zhang, Zheng, Hou, Shi, Fu, Duan, Huu, Wen, Zhang, Zhang, Li,
  Wang, Singer, Sun, Xu, Tafreshi, Lewis, Xia, and Zhang]{Kermany2018-ti}
Daniel~S Kermany, Michael Goldbaum, Wenjia Cai, Carolina C~S Valentim, Huiying
  Liang, Sally~L Baxter, Alex McKeown, Ge~Yang, Xiaokang Wu, Fangbing Yan,
  Justin Dong, Made~K Prasadha, Jacqueline Pei, Magdalene Y~L Ting, Jie Zhu,
  Christina Li, Sierra Hewett, Jason Dong, Ian Ziyar, Alexander Shi, Runze
  Zhang, Lianghong Zheng, Rui Hou, William Shi, Xin Fu, Yaou Duan, Viet A~N
  Huu, Cindy Wen, Edward~D Zhang, Charlotte~L Zhang, Oulan Li, Xiaobo Wang,
  Michael~A Singer, Xiaodong Sun, Jie Xu, Ali Tafreshi, M~Anthony Lewis, Huimin
  Xia, and Kang Zhang.
\newblock Identifying medical diagnoses and treatable diseases by image-based
  deep learning.
\newblock \emph{Cell}, 172\penalty0 (5):\penalty0 1122--1131.e9, February 2018.

\bibitem[Khan et~al.(2024)Khan, Akbar, Soomro, Hussain, Khalil, Khan, and
  Salam]{Khan2024-pm}
Irshad Khan, Wajahat Akbar, Abdullah Soomro, Tariq Hussain, Irshad Khalil,
  Muhammad~Nawaz Khan, and Abdu Salam.
\newblock Enhancing ocular health precision: Cataract detection using fundus
  images and {ResNet-50}.
\newblock \emph{IECE Transactions on Intelligent Systematics}, 1\penalty0
  (3):\penalty0 145--160, October 2024.

\bibitem[Koch et~al.(2015)Koch, Zemel, and Salakhutdinov]{koch2015siamese}
Gregory Koch, Richard Zemel, and Ruslan Salakhutdinov.
\newblock Siamese neural networks for one-shot image recognition.
\newblock In \emph{ICML Deep Learning Workshop}, 2015.

\bibitem[Koseoglu et~al.(2023)Koseoglu, Grzybowski, and Liu]{Koseoglu2023-ms}
Neslihan~Dilruba Koseoglu, Andrzej Grzybowski, and T~Y~Alvin Liu.
\newblock Deep learning applications to classification and detection of
  age-related macular degeneration on optical coherence tomography imaging: A
  review.
\newblock \emph{Ophthalmol. Ther.}, 12\penalty0 (5):\penalty0 2347--2359,
  October 2023.

\bibitem[Larxel(2020)]{larxel2022odir5k}
Larxel.
\newblock Ocular disease intelligent recognition (odir‑5k).
\newblock
  \url{https://www.kaggle.com/datasets/andrewmvd/ocular-disease-recognition-odir5k},
  2020.
\newblock Accessed: 2025-08-07.

\bibitem[LeCun et~al.(2015)LeCun, Bengio, and Hinton]{lecun2015deep}
Yann LeCun, Yoshua Bengio, and Geoffrey Hinton.
\newblock Deep learning.
\newblock \emph{Nature}, 521\penalty0 (7553):\penalty0 436--444, 2015.

\bibitem[Leng et~al.(2023)Leng, Shi, Wu, Zhu, Cai, Lu, and Liu]{Leng2023-ws}
Xiangjie Leng, Ruijie Shi, Yanxia Wu, Shiyin Zhu, Xingcan Cai, Xuejing Lu, and
  Ruobing Liu.
\newblock Deep learning for detection of age-related macular degeneration: A
  systematic review and meta-analysis of diagnostic test accuracy studies.
\newblock \emph{PLoS One}, 18\penalty0 (4):\penalty0 e0284060, April 2023.

\bibitem[Ling et~al.(2025)Ling, Chen, Yeh, Cheng, Huang, Shen, and
  Lee]{Ling2025-qh}
Xiao~Chun Ling, Henry Shen-Lih Chen, Po-Han Yeh, Yu-Chun Cheng, Chu-Yen Huang,
  Su-Chin Shen, and Yung-Sung Lee.
\newblock Deep learning in glaucoma detection and progression prediction: A
  systematic review and meta-analysis.
\newblock \emph{Biomedicines}, 13\penalty0 (2), February 2025.

\bibitem[Liu et~al.(2021)Liu, Lin, Cao, Hu, Wei, Zhang, Lin, and
  Guo]{liu2021swintransformerhierarchicalvision}
Ze~Liu, Yutong Lin, Yue Cao, Han Hu, Yixuan Wei, Zheng Zhang, Stephen Lin, and
  Baining Guo.
\newblock Swin transformer: Hierarchical vision transformer using shifted
  windows, 2021.
\newblock URL \url{https://arxiv.org/abs/2103.14030}.

\bibitem[Padalia et~al.(2022)Padalia, Mazumdar, and
  Singh]{padalia2022cnnlstmcombinationnetworkcataract}
Dishant Padalia, Abhishek Mazumdar, and Bharati Singh.
\newblock A cnn-lstm combination network for cataract detection using eye
  fundus images, 2022.
\newblock URL \url{https://arxiv.org/abs/2210.16093}.

\bibitem[Selvaraju et~al.(2017)Selvaraju, Cogswell, Das, Vedantam, Parikh, and
  Batra]{selvaraju2017grad}
Ramprasaath~R Selvaraju, Michael Cogswell, Abhishek Das, Ramakrishna Vedantam,
  Devi Parikh, and Dhruv Batra.
\newblock Grad-cam: Visual explanations from deep networks via gradient-based
  localization.
\newblock In \emph{Proceedings of the IEEE international conference on computer
  vision}, pages 618--626, 2017.

\bibitem[Selvaraju et~al.(2019)Selvaraju, Cogswell, Das, Vedantam, Parikh, and
  Batra]{Selvaraju_2019}
Ramprasaath~R. Selvaraju, Michael Cogswell, Abhishek Das, Ramakrishna Vedantam,
  Devi Parikh, and Dhruv Batra.
\newblock Grad-cam: Visual explanations from deep networks via gradient-based
  localization.
\newblock \emph{International Journal of Computer Vision}, 128\penalty0
  (2):\penalty0 336–359, October 2019.
\newblock ISSN 1573-1405.
\newblock \doi{10.1007/s11263-019-01228-7}.
\newblock URL \url{http://dx.doi.org/10.1007/s11263-019-01228-7}.

\bibitem[Sheng et~al.(2022)Sheng, Chen, Li, Ma, Yang, Bi, and
  Zhang]{Sheng2022-jd}
Bin Sheng, Xiaosi Chen, Tingyao Li, Tianxing Ma, Yang Yang, Lei Bi, and Xinyuan
  Zhang.
\newblock An overview of artificial intelligence in diabetic retinopathy and
  other ocular diseases.
\newblock \emph{Front. Public Health}, 10:\penalty0 971943, October 2022.

\bibitem[Steinmetz et~al.(2021)Steinmetz, Bourne, Briant, Flaxman, Taylor,
  Jonas, Abdoli, Abrha, Abualhasan, Abu-Gharbieh, et~al.]{steinmetz2021causes}
Jaimie~D Steinmetz, Rupert~RA Bourne, Paul~Svitil Briant, Seth~R Flaxman,
  Hugh~RB Taylor, Jost~B Jonas, Amir~Aberhe Abdoli, Woldu~Aberhe Abrha, Ahmed
  Abualhasan, Eman~Girum Abu-Gharbieh, et~al.
\newblock Causes of blindness and vision impairment in 2020 and trends over 30
  years, and prevalence of avoidable blindness in relation to vision 2020: the
  right to sight: an analysis for the global burden of disease study.
\newblock \emph{The Lancet Global Health}, 9\penalty0 (2):\penalty0 e144--e160,
  2021.

\bibitem[Tan and Le(2020)]{tan2020efficientnetrethinkingmodelscaling}
Mingxing Tan and Quoc~V. Le.
\newblock Efficientnet: Rethinking model scaling for convolutional neural
  networks, 2020.
\newblock URL \url{https://arxiv.org/abs/1905.11946}.

\bibitem[Ting et~al.(2017)Ting, Cheung, Lim, Tan, Quang, Gan, Hamzah,
  Garcia-Franco, San~Yeo, Lee, Wong, Sabanayagam, Baskaran, Ibrahim, Tan,
  Finkelstein, Lamoureux, Wong, Bressler, Sivaprasad, Varma, Jonas, He, Cheng,
  Cheung, Aung, Hsu, Lee, and Wong]{Ting2017-te}
Daniel Shu~Wei Ting, Carol Yim-Lui Cheung, Gilbert Lim, Gavin Siew~Wei Tan,
  Nguyen~D Quang, Alfred Gan, Haslina Hamzah, Renata Garcia-Franco, Ian~Yew
  San~Yeo, Shu~Yen Lee, Edmund Yick~Mun Wong, Charumathi Sabanayagam, Mani
  Baskaran, Farah Ibrahim, Ngiap~Chuan Tan, Eric~A Finkelstein, Ecosse~L
  Lamoureux, Ian~Y Wong, Neil~M Bressler, Sobha Sivaprasad, Rohit Varma, Jost~B
  Jonas, Ming~Guang He, Ching-Yu Cheng, Gemmy Chui~Ming Cheung, Tin Aung, Wynne
  Hsu, Mong~Li Lee, and Tien~Yin Wong.
\newblock Development and validation of a deep learning system for diabetic
  retinopathy and related eye diseases using retinal images from multiethnic
  populations with diabetes.
\newblock \emph{JAMA}, 318\penalty0 (22):\penalty0 2211, December 2017.

\bibitem[Touvron et~al.(2021)Touvron, Cord, Douze, Massa, Sablayrolles, and
  Jégou]{touvron2021trainingdataefficientimagetransformers}
Hugo Touvron, Matthieu Cord, Matthijs Douze, Francisco Massa, Alexandre
  Sablayrolles, and Hervé Jégou.
\newblock Training data-efficient image transformers \& distillation through
  attention, 2021.
\newblock URL \url{https://arxiv.org/abs/2012.12877}.

\bibitem[Tsiknakis et~al.(2021)Tsiknakis, Theodoropoulos, Manikis, Ktistakis,
  Boutsora, Berto, Scarpa, Scarpa, Fotiadis, and Marias]{Tsiknakis2021-oc}
Nikos Tsiknakis, Dimitris Theodoropoulos, Georgios Manikis, Emmanouil
  Ktistakis, Ourania Boutsora, Alexa Berto, Fabio Scarpa, Alberto Scarpa,
  Dimitrios~I Fotiadis, and Kostas Marias.
\newblock Deep learning for diabetic retinopathy detection and classification
  based on fundus images: A review.
\newblock \emph{Comput. Biol. Med.}, 135\penalty0 (104599):\penalty0 104599,
  August 2021.

\bibitem[Xie et~al.(2023{\natexlab{a}})Xie, Li, Wu, Zhao, Lin, Wang, Wang, Gu,
  Wang, Zheng, Jiang, and Chen]{Xie2023-ha}
He~Xie, Zhongwen Li, Chengchao Wu, Yitian Zhao, Chengmin Lin, Zhouqian Wang,
  Chenxi Wang, Qinyi Gu, Minye Wang, Qinxiang Zheng, Jiewei Jiang, and Wei
  Chen.
\newblock Deep learning for detecting visually impaired cataracts using fundus
  images.
\newblock \emph{Front. Cell Dev. Biol.}, 11:\penalty0 1197239, July
  2023{\natexlab{a}}.

\bibitem[Xie et~al.(2023{\natexlab{b}})Xie, Li, Wu, Zhao, Lin, Wang, Wang, Gu,
  Wang, Zheng, Jiang, and Chen]{Xie2023-kb}
He~Xie, Zhongwen Li, Chengchao Wu, Yitian Zhao, Chengmin Lin, Zhouqian Wang,
  Chenxi Wang, Qinyi Gu, Minye Wang, Qinxiang Zheng, Jiewei Jiang, and Wei
  Chen.
\newblock Deep learning for detecting visually impaired cataracts using fundus
  images.
\newblock \emph{Front. Cell Dev. Biol.}, 11:\penalty0 1197239, July
  2023{\natexlab{b}}.

\bibitem[Zhang et~al.(2022)Zhang, Hu, Xiao, Fang, Higashita, and
  Liu]{zhang2022machinelearningcataractclassification}
Xiaoqing Zhang, Yan Hu, Zunjie Xiao, Jiansheng Fang, Risa Higashita, and Jiang
  Liu.
\newblock Machine learning for cataract classification and grading on
  ophthalmic imaging modalities: A survey, 2022.
\newblock URL \url{https://arxiv.org/abs/2012.04830}.

\end{thebibliography}

\end{document}